# Optimal pooling strategies for laboratory testing

Brian G. Williams

South African Centre for Epidemiological Modelling and Analysis (SACEMA), Stellenbosch, South Africa

Correspondence to: Brian Williams at BrianGerardWilliams@gmail.com.

## Abstract
We consider the optimal strategy for laboratory testing of biological samples when we wish to know the results for each sample rather than the average prevalence of positive samples. If the proportion of positive samples is low considerable resources may be devoted to testing samples most of which are negative. An attractive strategy is to pool samples. If the pooled samples test positive one must then test the individual samples, otherwise they can all be assumed to be negative. The pool should be big enough to reduce the number of tests but not so big that the pooled samples are almost all positive.

We show that if the prevalence of positive samples is greater than 30% it is never worth pooling. From 30% down to 1% pools of size 4 are close to optimal. Below 1% substantial gains can be made by pooling, especially if the samples are pooled twice. However, with large pools the sensitivity of the test will fall correspondingly and this must be taken into consideration. We derive simple expressions for the optimal pool size and for the corresponding proportion of samples tested.

## Introduction

It is not uncommon for a large number of laboratory samples to be tested for signs of an infectious disease and this has been discussed for many infections including, for example, *Chlamydia trachomatis*,[1,2] HIV,[3-5] hepatitis B,[6] human papilloma virus,[7] and avian pneumovirus.[8] One may pool samples either to obtain the best estimate of the overall prevalence or to identify the infected individual, in both cases for the least cost. Here we are concerned to identify infected individuals. The purpose of this study is to develop approximate, but simple, formulae for determining the optimal pool size and the corresponding reduction in the number of tests that need to be done.

## Results

Let the sample size be $s_0$, the pool size be $s_1$ and the number of pools be $n_1$ so that $s_0 = s_1 n_1$. We first do one test on each pool. If $p_0$ is the prevalence in individual samples the prevalence in the pooled samples will be

$$p_1 = 1 - (1 - p_0)^{s_1} \qquad 1$$

and for this proportion of the pooled samples $s_1$ further tests will be carried out to determine the individual results. The expected number of tests, $T$, is therefore

$$E(T) = n_1 + s_0 p_1 \approx s_0 \left( \frac{1}{s_1} + p_0 s_1 \right) \qquad 2$$

provided $p_0 s_1 \ll 1$. To determine the optimal pool size we wish to determine $\tilde{s}_1$, the value of $s_1$ which minimises this expression. Since

$$\frac{dE(T)}{ds_1} \approx s_0 \left( -\frac{1}{s_1^2} + p_0 \right) \qquad 3$$

the minimum value is obtained when

$$\tilde{s}_1 \approx \frac{1}{\sqrt{p_0}}. \qquad 4$$

and with Equation 2 the proportion tested is approximately $2p_0^{0.5}$.[1] Keeping the next term in Equation 2 shows that the fractional error in this approximation is of the order $p_0/4$. Since we are mainly interested in small values of $p_0$ the approximation is generally good. Table 1 gives the approximate and exact estimates of the optimal pool size and the number of tests as a proportion of the number of samples for a range of values of the prevalence of the infection. As expected, the optimal pool size decreases, and the number of tests increases, as the prevalence increases. However, the approximation is good up to a prevalence of 10% and even at 25% the approximate result suggests that there is no gain while the exact result shows that the number of tests would be reduced by 9%.

Table 1. Approximate and exact estimates of the optimal pool size, $s_1$, and the proportion tested $T$ for different values of the prevalence, $p_0$.

| $p_0$ (%) | Approximate | | Exact | |
| --- | --- | --- | --- | --- |
| | $s_1$ | $T$ (%) | $s_1$ | $T$ (%) |
| 0.01 | 100.0 | 2.0 | 100.5 | 2.0 |
| 0.1 | 31.6 | 6.3 | 32.1 | 6.3 |
| 1.0 | 10.0 | 20.0 | 10.5 | 19.5 |
| 5.0 | 4.5 | 44.7 | 5.0 | 42.6 |
| 10.0 | 3.2 | 63.2 | 3.8 | 59.3 |
| 25.0 | 2.0 | 100.0 | 2.8 | 91.0 |

Assuming that pooling is not worth doing unless we can reduce the total number of tests to be done by at least 25% requires that $E(T)$, the expected number of tests, must be less than 75% of the sample size, $s_1$. From Equations 1 and 2 it is not worth pooling at a size greater than $s$ if the prevalence is greater than

$$\tilde{p}_0 = 1 - \left(\frac{1}{s_1} + 0.25\right)^{1/s_1} \quad 5$$

---

[1] If we calculate the exact values and fit a power-law curve to the data for values of the prevalence between 0.001% and 10% the proportion tested is $1.5 p_0^{0.46}$.

As shown in Table 2 it is not worth pooling if the prevalence is greater than 33%. It is worth pooling in groups of 2 or more at any prevalence below 33%. It is not worth pooling in groups of 16, for example, unless the prevalence is less than 16%.

Table 2. The critical prevalence at which the number of tests carried out will be 25% less than the number of samples collected) for different pool sizes.

| Pool size | 2 | 4 | 8 | 16 | 32 | 64 | 128 | 256 |
|---|---|---|---|---|---|---|---|---|
| Critical prev. (%) | 33 | 30 | 23 | 16 | 10 | 6.3 | 3.7 | 2.1 |

**Pooling pools**
If the prevalence is low then the pooled prevalence is low and this raises the possibility of pooling the pools. For example, one might group samples in pools of sixteen. The samples in any positive pool would then be tested in pools of four, say, and any of these that are positive would be tested individually.

We therefore extend the logic of the previous calculations. To keep the notation clear we will start (nominally) with a single group so that $n_0 = 1$ containing $s_0$ samples so that $s_0$ is the sample size. We divide the samples into $n_1$ groups each containing $s_1$ samples, then into $n_2$ groups each containing $s_2$ samples and finally into $n_3$ groups each containing $s_3$ samples. Then $n_3 = s_0$ and $s_3 = 1$. We initially test all samples in the first set of pools, then a proportion $p_1$ in the second set of pools, and then a proportion $p_1 p_2$ individually. The number of tests that we do is

$$E(T) = n_1 + n_2 p_1 + n_3 p_1 p_2 \quad 6$$

To calculate $p_1$ we use Equation 1, However, to calculate $p_2$ we need to allow for the fact that by excluding any first level pools that are negative we have changed the denominator for the calculation of the prevalence. We therefore divide the expression for $p_2$ by $1 - p_1$.

We again get a useful approximation to the optimal number of pools at each level by considering what happens if we pool once, as above, and then treat this as new pooling problem and pool again. Then the proportion of samples that are tested is approximately $4p_0^{0.5} p_1^{0.5}$ and since $p_1 \approx p_0 s_1 \approx p_0^{0.5}$ the proportion tested is $4p_0^{0.75}$.[2]

---
[2] If we calculate the exact values and fit a power-law curve to the data for values of the prevalence between 0.001% and 10% the proportion tested is then given by $3.3p_0^{0.77}$.

Table 3. The percent prevalence in the original samples ($p_0$), the optimal pool size for the first pool ($s_1$), the second set of pools ($s_2$) and the number tested as a percentage of the number of samples using the approximate and exact formulae.

| | Approximate | | | Exact | | |
|---|---|---|---|---|---|---|
| $p_0$ (%) | $s_1$ | $s_2$ | $T$ (%) | $s_1$ | $s_2$ | $T$ (%) |
| 0.01 | 100.0 | 10.0 | 0.4 | 87.6 | 7.3 | 0.3 |
| 0.1 | 31.6 | 5.6 | 2.2 | 30.2 | 6.6 | 1.6 |
| 1.0 | 10.0 | 3.2 | 12.6 | 9.4 | 5.1 | 8.9 |
| 5.0 | 4.5 | 2.1 | 42.3 | 4.4 | 4.2 | 29.8 |

Table 3 shows that the results of the approximate and exact formulae are close up to a prevalence of 1%. When the prevalence is very low, pooling twice can lead to substantial savings compared to pooling once. Comparing the data in Table 1 and Table 3 we see that at a prevalence of 0.01% pooling twice can reduce the number of tests that are done from 2% of the sample size to 0.4% of the sample size.

## Conclusion

The optimal pool-size for laboratory tests is given, with reasonable accuracy, by $p_0^{-0.5}$, one over the square root of the prevalence, and the number of tests that then need to be done as a proportion of the number of samples is given by $2p_0^{0.5}$, twice the square root of the prevalence. If the pool sizes found in this way are large, then one can take each pooled sample, calculate the expected prevalence in the pooled samples as $p_0 s_1$ and calculate the size of the next level of pools as before. The number of tests that then need to be done is approximately $4p_0^{0.75}$

A practical recommendation is that one should never pool samples above a prevalence of about 30%. From 30% down to 1% pools of size 4 are close to optimal. Below 1% substantial gains can be made by pooling especially if the samples are pooled twice. However, with large pools the sensitivity of the test will fall correspondingly and this must be considered. If the test is extremely sensitive and there are no constraints arising from sample preparation and management, it is likely to be worth pooling samples in very large batches.

## Discussion

The approximate expressions noted in the previous sections provide a simple way of estimating the optimal size of pools and the gains that will be had from doing this. We illustrate the importance of this with reference to several published studies.

In a study of the incidence of HIV, measurements of the p24 antigen were used as a marker of recent HIV infection.[3] There were 700 samples and the prevalence was 1.1%. The authors pooled their samples into pools of size 100, then of size 50, then of size 10, and finally tested individual samples. This led them to conduct 150 tests, a considerable saving over 700. However, the results presented here suggest that they could have grouped samples into pools of size 9 and then 3. The expected number of tests would then have been 122. While the number of tests that the authors carried out was close to optimal, the latter pooling strategy (pool sizes: 9, 3, 1) would have been much easier to implement than the strategy that was used (pool sizes: 100, 50, 10, 1). Furthermore, as the pool size increases there is bound to be a loss in sensitivity especially when the pool size is as large as 100.

In a study of *Chlamydia trachomatis*[2] the prevalence of infection was 4.5%, samples were pooled in batches of 5, and this led to a saving of 60% in the number of tests that were done. Using the expressions derived here the optimal pool size is 4.7, so that their pool size was optimal, and the expected number of tests is 42% of the sample size, in close agreement with their result.

In a study of acute HIV-infection in California[9] the authors note that the San Francisco Public Health Laboratory initially used pools of size 50 and then 10 but later switched to pooling once in pools of size 10. The measured prevalence was 0.36%. The Los Angeles County Public Health Laboratory used pools of size 90 and then 10. The measured prevalence was 0.06%. This analysis suggests that the optimal pooling scheme would have been to use polls of size 16 and 4 in the first case and 42 and 6 in the second case. This would have reduced the number of tests to 4% and 1% of the number of samples in the two cases.

There are two practical issues which may affect the optimal pooling strategy. First, if one is pooling samples the amount of each of the original samples in the pooled sample will be less than would be the case if the samples were not pooled. Since this might affect the sensitivity of the test it is important to ensure that the pooling does not significantly reduce the sensitivity. For example, the detection limit for HIV in plasma is typically when the concentration falls to below about 100/μL. Polling samples in groups of 100, say, would raise this to about 10,000/μL which would be of little use. This problem is alluded to in the discussion of HIV testing in California.[9]

Secondly, one has to consider the practicalities of pooling. If raw (blood or urine) samples can be pooled before any chemical analysis is done then the above arguments apply. Suppose, however, that half of the work and cost involves steps which need to be carried out before the pooling is done. Then no matter what size the pools one could never reduce the amount of work or the costs by more than 50%. Furthermore the practicalities of pooling (avoiding contamination and so on) might mean that unless the benefit is very substantial the pooling is not worth doing.